\documentclass[aps,nofootinbib,superscriptaddress, showpacs,preprintnumbers,  nofootinbibt,twocolumn]{revtex4-2}
\usepackage{epsfig}
\usepackage{multirow}
\usepackage{subcaption}
\usepackage{eurosym}
\usepackage{dcolumn}
\usepackage{bm}
\usepackage{enumerate}
\usepackage{float}
\usepackage{epstopdf}
\usepackage{amsmath}
\usepackage{bm}
\usepackage{amsfonts}
\usepackage{amssymb}
\usepackage{graphicx}
\usepackage{alphalph,mathtools}
\usepackage{etoolbox}
\usepackage{color}
\usepackage{booktabs}
\usepackage{hyperref}
\hypersetup{colorlinks,citecolor=blue}
\usepackage{footnote}
\usepackage{makecell,tabularx}

\def\be{\begin{equation}}
\def\ee{\end{equation}}
\def\bea{\begin{eqnarray}}
\def\eea{\end{eqnarray}}

\begin{document}

\title{\bf Observational Constraints on the Parameters of Hořava-Lifshitz Gravity}
\author{Himanshu Chaudhary}
\email{himanshuch1729@gmail.com}
\affiliation{Department of Applied Mathematics, Delhi Technological University, Delhi-110042, India}
\affiliation{Pacif Institute of Cosmology
and Selfology (PICS), Sagara, Sambalpur 768224, Odisha, India} 
\author{Ujjal Debnath}
\email{ujjaldebnath@gmail.com} 
\affiliation{Department of
Mathematics, Indian Institute of Engineering Science and
Technology, Shibpur, Howrah-711 103, India}
\author{Shibesh Kumar Jas Pacif}
\email{shibesh.math@gmail.com}
\affiliation{Pacif Institute of Cosmology
and Selfology (PICS), Sagara, Sambalpur 768224, Odisha, India} 
\author{Niyaz Uddin Molla}
\email{niyazuddin182@gmail.com}
\affiliation{Department of
Mathematics, Indian Institute of Engineering Science and Technology, Shibpur, Howrah-711 103, India}
\author{G.Mustafa}
\email{gmustafa3828@gmail.com} 
\affiliation{Department of Physics,
Zhejiang Normal University, Jinhua 321004, Peoples Republic of China}
\affiliation{New Uzbekistan University, Mustaqillik ave. 54, 100007 Tashkent, Uzbekistan}
\author{S. K. Maurya}
\email{sunil@unizwa.edu.om} \affiliation{Department of Mathematical and Physical Sciences, College of Arts and Sciences, University of Nizwa, Nizwa 616, Sultanate of Oman.}

\begin{abstract}
This study investigates the accelerated cosmic expansion within the Hořava-Lifshitz Model. To constrain the cosmological parameters of this model, we incorporate 17 Baryon Acoustic Oscillation points, 31 Cosmic Chronometer points, 40 Type Ia Supernovae points, 24 quasar Hubble diagram points, and 162 Gamma Ray Bursts points, along with the latest Hubble constant measurement (R22). We treat \(r_{d}\) as a free parameter to extract \(H_{0}\) and \(r_{d}\) using late-time datasets, aiming for optimal fitting values in each model. Treating \(r_{d}\) as free improves precision, reduces bias, and enhances dataset compatibility. The obtained values of \(H_{0}\) and \(r_{d}\) are compared to the \(\Lambda\)CDM model, showing consistency with previous estimates from Planck and SDSS studies. The Akaike Information Criterion (AIC) and Bayesian Information Criterion (BIC) favor the Hořava-Lifshitz model, with the \(\Lambda\)CDM model having the lowest AIC. {Additionally, we conduct \(\Delta\)AIC and \(\Delta\)BIC analyses to assess model preference.} Validation using the reduced \(\chi_{red}^{2}\) statistic indicates satisfactory fits for the Hořava-Lifshitz model, while recognizing \(\Lambda\)CDM as the preferred model. Extensions of the analysis warrant further investigation.
\end{abstract}
\maketitle

\section{Introduction}
Einstein's General theory of  relativity (GR) remains the current gold standard for gravitational theories, consistently passing all tests to date \cite{will2018theory,Berti:2015itd,will2014confrontation}. Nevertheless, it is important to develop new theories to further test GR and to explore the potential for a theory of gravity beyond GR. Current research interests include finding a valid theory of quantum gravity and explaining cosmological phenomena such as dark energy and dark matter\cite{jain2010cosmological,clifton2012modified,koyama2016cosmological}. One challenge with GR is that it is not power-counting renormalizable. To address this, Hořava proposed a theory of gravity beyond GR that is both renormalizable and ultraviolet complete \cite{Horava:2009uw}. This theory, known as Hořava-Lifshitz (HL) gravity, breaks Lorentz invariance in the ultraviolet regime by introducing a Lifshitz-type anisotropic scaling between space and time.\\\\
In Einstein's gravitational theory, a prominent challenge arises from ultraviolet (UV) divergence, known as the UV completion problem. Stelle (1976) addressed this by proposing the incorporation of higher-order curvature invariants into gravitational theories \cite{14}. However, this approach encountered difficulties due to the appearance of ghost degrees of freedom associated with higher-order derivatives. Horava (2009) effectively tackled this issue by introducing additional terms featuring higher-order spatial curvature components into the action, leading to what is now termed deformed HL gravity \cite{15}.  Under high-energy conditions, HL gravity is formulated by abandoning Lorentz symmetry through a Lifshitz-type rescaling process \cite{16}. This rescaling involves altering the scaling properties of space and time coordinates, characterized by an exponent denoted as $z$, and $d$ representing the spacetime dimension. Lorentz symmetry is broken when $z\neq1$ and restored only when $z=1$. Modified HL gravity has garnered significant attention in scientific literature \cite{17,18,19,20,21,22,23}. Researchers have explored various aspects, including cosmological implications, such as the Friedmann equation within the deformed HL gravity framework for the FLRW Universe \cite{24}, novel approaches to dark energy like holographic dark energy (HDE) \cite{25}, and investigations into black hole properties such as quasi-normal modes \cite{26}. Additionally, studies have delved into models involving ghost dark energy within the context of deformed HL gravity \cite{27}. The HL gravity model has been extensively examined by numerous researchers \cite{chaichian2010modified,carloni2010modified,elizalde2010unifying,nojiri2010covariant,nojiri2010proposal,nojiri2011covariant,bamba2012dark}., exploring its diverse implications and applications across cosmology and gravitational physics.\\\\
HL gravity represents a significant endeavor in theoretical physics, aiming to address the quantum gravity problem and rectify the lack of renormalizability inherent in standard general relativity. By providing a UV-complete theory, it seeks to offer a framework capable of describing gravitational interactions at both small quantum scales and large cosmological scales. Despite its ambitious goals, the theory has not been without challenges and criticisms. One major point of contention revolves around its viability as a quantum theory. Critics have raised concerns about the presence of ghost instabilities, which can lead to inconsistencies and unpredictability within the theory's predictions. These challenges have spurred ongoing debate within the scientific community regarding the feasibility and robustness of HL gravity as a fundamental theory of gravity. Nevertheless, researchers remain engaged in studying and refining the HL gravity framework, particularly in its implications for cosmology and black hole physics. Understanding the cosmological consequences of this theory is of particular interest, as it may offer insights into the nature of the universe at large scales. Investigations into black hole properties within this framework also hold promise for elucidating fundamental aspects of gravitational physics. It's important to recognize that the scientific community's understanding of HL gravity is dynamic and subject to change as further research and experimental data become available. Jawad et al. \cite{28}  investigated the dynamical stability and cosmographic characteristics of some cosmic models under deformed HL gravity, matching their results with observational data. They studied cosmographic parameters and thermodynamics with regard to Einstein's gravity and deformed HL gravity associated with Kaniadakis HDE~\cite{29}. Additionally, they explored the cosmological consequences of Sharma-Mittal holographic dark energy (SMHDE), analysing cosmological parameters and thermodynamics~\cite{dubey2021sharma}. {Further, constraining a model with observations is crucial for ensuring its physical relevance and predictive power. Observational constraints help in refining theoretical models by aligning them with empirical data, thus validating or refuting theoretical predictions. Following the theoretical formulation of HL gravity, various datasets were used by \cite{dutta2010observational} to constrain the corresponding cosmological scenarios. Additionally, \cite{dutta2010overall} focused on the running parameter \(\lambda\) of HL gravity, which governs the flow between the Ultra-Violet and the Infra-Red. They found that \(\lambda\) is restricted to \(|\lambda - 1| \lesssim 0.02\), with the best fit value being \(|\lambda_{\text{b.f.}} - 1| \approx 0.002\). This observational analysis confines the running parameter \(\lambda\) very close to its IR value of 1. It is important to note that basic versions of HL gravity may encounter perturbative instabilities, as highlighted in previous studies \cite{bogdanos2010perturbative,charmousis2009strong}. These instabilities pose challenges to the theory's robustness and reliability. Recognizing and addressing these issues is essential for advancing HL gravity and ensuring its consistency with empirical observations and theoretical expectations.}\\\\
Motivated by the preceding discourse, this study explores into the realm of HL gravity, which emerges as a compelling alternative to Einstein's gravitational theory for several reasons. Both theories aim to describe the behavior of gravity, but they approach the challenge from different perspectives and with different goals in mind. One key motivation for considering HL gravity over Einstein gravity lies in addressing the UV completion problem, a significant challenge in theoretical physics. HL gravity offers a promising avenue for resolving UV divergences by introducing higher-order spatial curvature components into the action, thereby providing a UV-complete theory capable of describing gravitational interactions at both small quantum scales and large cosmological scales.
Our primary objective is to meticulously evaluate the compatibility of HL model with observational data, thus deriving rigorous constraints on crucial cosmic parameters.\\\\ 
Harnessing recent measurements of the Hubble parameter, \(H(z)\), extracted from cosmic chronometers, alongside a diverse array of datasets spanning Type Ia Supernovae, Gamma-Ray Bursts (GRB), Quasars, and uncorrelated Baryon Acoustic Oscillations (BAO), our analysis offers a comprehensive survey of the cosmic terrain. Additionally, we aim to probe the ongoing cosmological tensions such as \(H_{0}\) and the sensitivity of \(r_{d}\). The paper is structured as follows: In Section \ref{sec1}, we present the basic equations of HL Gravity. In Section \ref{sec3}, we focus on the methodology employed to constrain the crucial parameters of HL model. Our study's outcomes are detailed in Section \ref{sec4}, where we present the results obtained. Finally, in Section \ref{sec5}, we conclude the paper, offering discussions on the implications and significance of our findings.\\\\
\section{Hořava-Lifshitz Gravity and Basic Equations}\label{sec1}
We assume the Arnowitt-Deser-Misner decomposition, which is provided as in the metric form:
\begin{equation}\label{HL1}
ds^2=g_{i j} \left(dx^i+N^i dt\right)\left(dx^j+N^j dt\right)-N^2dt^2.
\end{equation}
In the above equation, $N$ is representing the lapse function, $N_i$ is providing the shift vector, $g_{ij}$ is denoting the spatial metric. For the present analysis, the scaling transformation in the framework of coordinates reads as $t\rightarrow l^3 t$ and $x^i\rightarrow l x^i$. The action of the HL gravity has two major components, namely, the
potential term and kinetic term. The HL gravity action within the scope of potential term and kinetic term is defined as:
$$S_g=S_v+S_k=\int dt d^3 x \sqrt{g} N\left(L_v+L_k \right),$$ where, the kinetic term is defined as
$$S_k=\int dt d^3 x \sqrt{g} N \left[\frac{2\left(K_{ij}K^{ij}
-\lambda K^2\right)}{\kappa^2}\right],$$
where, The extrinsic curvature is provided as 
$$K_{ij}=\frac{\dot{g}_{ij}-\Delta_i N_j-\Delta_j N_i}{2N}.$$
The number of invariants, while working with the Lagrangian, $L_v$, can be reduced due to its symmetric property. This symmetry is referred to as detailed balance. BY using this detailed balance, the action can be revised as
\begin{widetext}
\begin{eqnarray}
S_g= \int dt d^3x \sqrt{g} N \left[\frac{2\left(K_{ij}K^{ij}
-\lambda K^2\right)}{\kappa^2}+\frac{\kappa^2
C_{ij}C^{ij}}{2\omega^4} -\frac{\kappa^2 \mu \epsilon^{i j k }
R_{i, j} \Delta_j R^l_k}{2\omega^2
\sqrt{g}}\right.\nonumber\\\left.+\frac{\kappa^2 \mu^2 R_{ij} R^{ij}}{8}
-\frac{\kappa^2
\mu^2}{8(3\lambda-1)}\left\{\frac{(1-4\lambda)R^2}{4} +\Lambda R
-3 \Lambda^2 \right\}\right],
\end{eqnarray}
 $$\text{where}~~~~~~C^{ij}=\frac{\epsilon^{ijk}
\Delta_k\left(R_i^j-\frac{R}{4} \delta^j_i\right)}{\sqrt{g}}.$$
\end{widetext}
is the known as the Cotton tensor. Here the covariant derivatives can be  determined by the spatial metric $g_{ij} \epsilon^{ijk}$, which is the antisymmetric unit tensor. Here, the parameter $\lambda$ is the dimensionless
constant and the other parameters $\kappa$, $\omega$ and $\mu$ are also the constants.\\\\
Horava obtained a gravitational action by considering 
the temporal dependency of the lapse function $N(t)$. For FRW metric with the assumptions $N=1~,~g_{ij}=a^2(t)\gamma_{ij}~,~N^i=0$ and
$$\gamma_{ij}dx^i dx^j=\frac{dr^2}{1-kr^2}+r^2 d\Omega_2^2,$$
and by taking the variations of $N$ and $g_{ij}$, we can obtain
the two Field equations which are given below:
\begin{widetext}
\begin{equation}\label{HLFriedmann1}
H^2=\frac{\kappa^2\rho_m}{6\left(3\lambda-1\right)}
+\frac{\kappa^2}{6\left(3\lambda-1\right)}\left[\frac{3\kappa^2\mu^2
k^2} {8\left(3\lambda-1\right)a^4}+\frac{3\kappa^2\mu^2 \Lambda^2}
{8\left(3\lambda-1\right)}\right]-\frac{\kappa^4 \mu^2 \Lambda
k}{8\left(3\lambda-1\right)^2a^2},
\end{equation}
\begin{equation}\label{HLFriedmann2}
\dot{H}+\frac{3H^2}{2}=-\frac{\kappa^2
p_m}{4\left(3\lambda-1\right)} -\frac{\kappa^2}
{4\left(3\lambda-1\right)}\left[\frac{3\kappa^2\mu^2 k^2}
{8\left(3\lambda-1\right)a^4}+\frac{3\kappa^2\mu^2 \Lambda^2}
{8\left(3\lambda-1\right)}\right]-\frac{\kappa^4 \mu^2 \Lambda
k}{8\left(3\lambda-1\right)^2a^2}.
\end{equation}
\end{widetext}
where $k=0,\pm 1$ represent the flat, closed and open  universe
respectively. In the field equations mentioned above, the term proportional to $\frac{1}{a^4}$ is treated as the "Dark radiation" term and $\Lambda$ denotes the cosmological constant. Also, $H = \frac{\dot{a}}{a}$ represents the Hubble parameter.
Now we assume that the universe is filled with dark matter and we have not taken any external dark energy because the modified gravity can produce dark energy. So the conservation equation of the dark matter (DM) is given by
\begin{equation}\label{DM}
    \dot{\rho}_m + 3H(\rho_m + p_m) = 0,
\end{equation}
Since the dark matter has negligible  pressure (i.e., $p_m \sim 0$), so the equation
\eqref{DM} yields $\rho_m = \rho_{m0}a^{-3}$.  
Using the dimensionless parameters $\Omega_{m0}\equiv\frac{\rho_{m0}}{3H_0^2}$, $\Omega_{k0}=-\frac{k}{H_0^2}$, 
$\Omega_{\Lambda 0}=\frac{\Lambda}{2H_0^2}$, we obtain (choosing $\kappa^2=1$)
\begin{widetext}
\begin{equation}\label{E}
H^2(z)=\frac{2H_0^2 \Omega_{m0}(1+z)^3}{3\lambda-1}+\frac{\mu^2 H_0^2} {16(3\lambda-1)^2}\left[\Omega_{k0} (1+z)^2
+3 \Omega_{\Lambda0} \right]^2
\end{equation}
\end{widetext}
\section{Methodology}\label{sec3}
In our investigation, we carefully picked a specific set of recent Baryon Acoustic Oscillation (BAO) measurements extracted from multiple galaxy surveys, with a primary focus on data from the Sloan Digital Sky Survey (SDSS) \cite{42BAO,43BAO,44BAO,45BAO,46BAO,47BAO}. To enrich our dataset, we augmented it with valuable contributions from other surveys such as the Dark Energy Survey (DES) \cite{48BAO}, the Dark Energy Camera Legacy Survey (DECaLS) \cite{49BAO}, and 6dFGS BAO \cite{50BAO}. Recognizing the potential for interdependencies among our selected data points, we diligently addressed this concern. We carefully curated a subset to mitigate highly correlated points, fully appreciating the importance of managing possible correlations within our chosen dataset. To ensure a comprehensive assessment of systematic errors, we employed mock datasets generated from N-body simulations to precisely calculate covariance matrices. Obtaining accurate covariance matrices among measurements derived from diverse observational surveys presented a significant challenge. To address a significant challenge, we utilized covariance analysis as outlined in reference \cite{38BAO}. Specifically, we adopted the approach where the diagonal elements of the covariance matrix, denoted as $C_{ii}$, were set equal to the square of the 1$\sigma$ errors, $\sigma_{i}^{2}$. In order to introduce correlations among our selected subset of data, we augmented the covariance matrix by incorporating non-diagonal elements while preserving its symmetry. This was achieved by randomly pairing data points, resulting in a total of twelve pairs. For each pair, the non-diagonal elements were determined using the expression $C_{ij} = 0.5 \sigma_{i} \sigma_{j}$, where $\sigma_i$ and $\sigma_j$ represent the 1$\sigma$ errors associated with data points $i$ and $j$, respectively. Through this methodology, we effectively simulated correlations within 56\% of the BAO dataset under consideration. This approach enabled us to capture the interdependence among data points, providing a more comprehensive representation of the underlying structure within the dataset. By incorporating these correlations, we aimed to enhance the accuracy and reliability of our analyses, thereby improving our understanding of the phenomenon under investigation. To constrain our cosmological model parameters, we expanded our dataset by including various additional measurements. Firstly, we incorporated thirty independent Hubble parameter measurements derived from cosmic chronometers (CC) detailed in \cite{CC1,CC2,CC3,CC4}. Additionally, our study integrates recent Pantheon Type Ia Supernovae samples \cite{smith2020first}, 24 binned quasar \cite{Quasar} distance modulus data, 162 Gamma-Ray Bursts (GRBs) \cite{GRB}, and the latest Hubble constant measurement (R22) as an additional prior \cite{7BAO}. To analyze these datasets, we adopted a nested sampling approach using the Polychord package \cite{PolyChord}, enabling thorough exploration of parameter space for determining optimal fit values. Additionally, we utilized the GetDist package \cite{Getdist} to present our findings clearly and informatively, enhancing the interpretation and visualization of the analysis outcomes. This comprehensive methodology ensures a robust understanding of cosmological parameters while providing insights into the Universe's fundamental properties.
\begin{figure*}[!htp]
\centering
\includegraphics[scale=0.72]{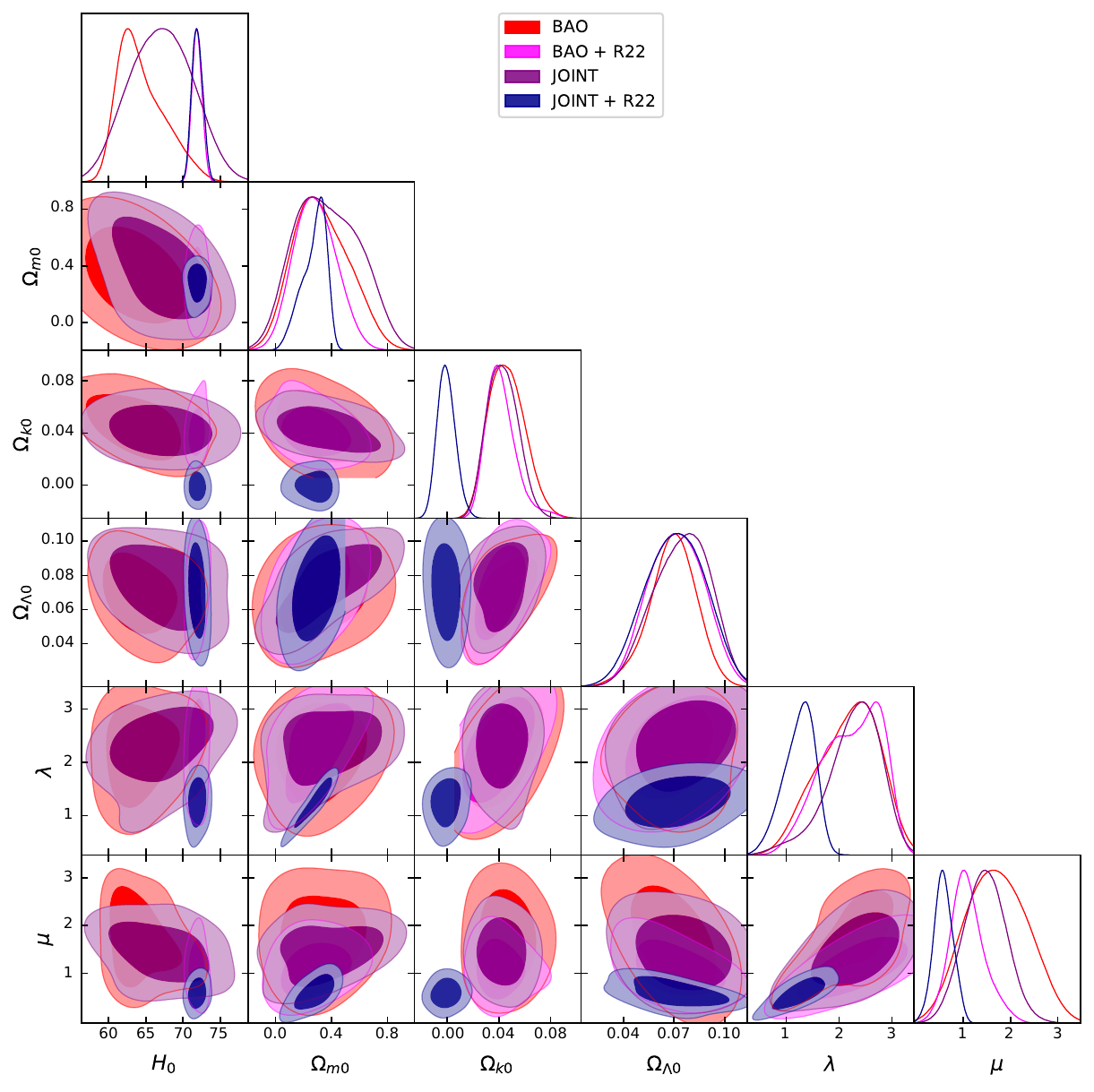}
\caption{The figure illustrates the posterior distribution of various observational data measurements using the HL Model, highlighting the 1$\sigma$ and 2$\sigma$ regions.}\label{fig_1}
\end{figure*}
\begin{figure}
\centering
\includegraphics[scale=0.42]{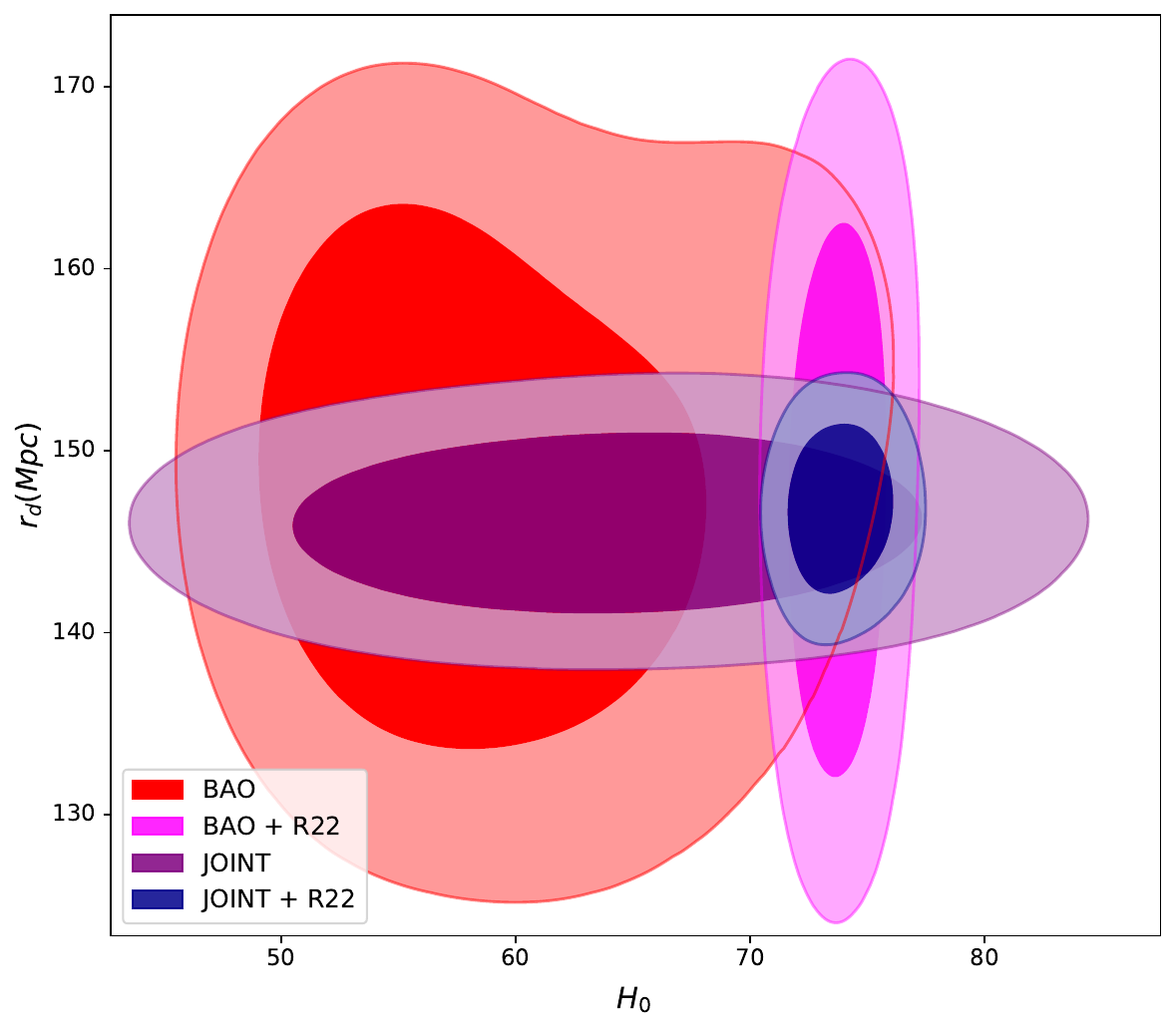}
\caption{The figure illustrates the posterior distribution of various observational data measurements using the HL Model, highlighting the 1$\sigma$ and 2$\sigma$ regions.}\label{fig_2}
\end{figure}
\begin{table*}
\centering
\begin{tabular}{|c|c|c|c|c|c|c|}
\hline
\multicolumn{7}{|c|}{MCMC Results} \\
\hline\hline
 Model & Parameter & Priors & BAO & BAO + R22 & Joint & Joint + R22 \\[1ex]
\hline
& $H_0$ & [50.,80.] & $70.256500_{\pm 5.502574}^{\pm 9.206346}$ & $73.798696_{\pm 1.315504}^{\pm 2.543240}$ & $69.837362_{\pm 1.204873}^{\pm 2.478047}$ & $71.508817_{\pm 0.882681}^{\pm 1.614354}$ \\[1ex]
$\Lambda$CDM Model &$\Omega_{m0}$ &[0.,1.]  & $0.270161_{\pm 0.013827}^{\pm 0.033280}$ & $0.268255_{\pm 0.016790}^{\pm 0.043787}$ & $0.274526_{\pm 0.009858}^{\pm 0.021367}$ & $0.270609_{\pm 0.009678}^{\pm 0.022243}$ \\[1ex]
&$r_{d}$(Mpc)  & [100.,200.] & $145.807497_{\pm 9.060325}^{\pm 13.994434}$ & $138.345762_{\pm 2.451777}^{\pm 4.939114}$ & $145.811932_{\pm 2.347736}^{\pm 4.545176}$ & $142.591327_{\pm 1.850175}^{\pm 3.951717}$ \\[1ex]
&$r_{d}/r_{fid}$& [0.9,1.1] & $0.970661_{\pm 0.055989}^{\pm 0.068973}$ & $0.930500_{\pm 0.021272}^{\pm 0.028238}$ & $0.974064_{\pm 0.033510}^{\pm 0.057975}$ & $0.947662_{\pm 0.028335}^{\pm 0.043021}$ \\[1ex]
\hline
& $H_{0}$ &  [50.,80.] & $65.700835_{\pm 6.213905}^{\pm 8.188232}$ & $73.764424_{\pm 1.206257}^{\pm 2.238343}$ & $69.083097_{\pm 9.523705}^{\pm 13.449031}$ & $73.841896_{\pm 1.470940}^{\pm 2.325034}$ \\[1ex]
& $\Omega_{m0}$ & [0.,1.] & $0.307520_{\pm 0.246232}^{\pm 0.374679}$ & $0.313934_{\pm 0.193706}^{\pm 0.279260}$ & $0.301674_{\pm 0.290568}^{\pm 0.441591}$ & $0.316901_{\pm 0.121592}^{\pm 0.251704}$ \\[1ex]
& $\Omega_{k0}$ & [0.,0.1] & $0.045949_{\pm 0.015391}^{\pm 0.019647}$ & $0.041633_{\pm 0.011413}^{\pm 0.015582}$ & $0.042779_{\pm 0.013332}^{\pm 0.017139}$ & $-0.000867_{\pm 0.006873}^{\pm 0.011261}$ \\[1ex]
Hořava-Lifshitz Model & $\Omega_{\Lambda0}$ & [0.,0.1] & $0.069321_{\pm 0.011309}^{\pm 0.031470}$ & $0.071208_{\pm 0.016735}^{\pm 0.025913}$ & $0.074617_{\pm 0.017924}^{\pm 0.032586}$ & $0.070843_{\pm 0.018968}^{\pm 0.032504}$ \\[1ex]
& $\lambda$ & [0.,1.] & $2.179575_{\pm 0.718167}^{\pm 1.092257}$ & $2.250670_{\pm 0.613784}^{\pm 1.174342}$ & $2.286921_{\pm 0.450757}^{\pm 1.402826}$ & $1.249560_{\pm 0.309072}^{\pm 0.687458}$ \\[1ex]
& $\mu$ & [0.,3.] & $1.709372_{\pm 0.696013}^{\pm 1.036271}$ & $1.090299_{\pm 0.313413}^{\pm 0.706884}$ & $1.457249_{\pm 0.445976}^{\pm 0.864455}$ & $0.581412_{\pm 0.201536}^{\pm 0.432033}$ \\[1ex]
&$r_{d}$ (Mpc) & [100.,200.] & $148.489236_{\pm 8.942271}^{\pm 14.876427}$ & $147.612337_{\pm 9.931756}^{\pm 13.663930}$ & $146.021893_{\pm 2.757663}^{\pm 5.394418}$ & $146.862925_{\pm 3.047961}^{\pm 5.019343}$ \\[1ex]
&$r_{d}/r_{fid}$ & [0.9,1.1] &$1.004965_{\pm 0.058313}^{\pm 0.099047}$ & $0.993782_{\pm 0.064095}^{\pm 0.086915}$ & $0.979637_{\pm 0.020247}^{\pm 0.032327}$ & $0.991474_{\pm 0.020670}^{\pm 0.034560}$ \\[1ex]
\hline
\end{tabular}
\caption{Summary of the MCMC Results.}
\label{tab_2}
\end{table*}
\section{Results}\label{sec4}
Fig ~\ref{fig_1} shows the \(68\%\) and \(95\%\) confidence levels for important cosmological parameters in the HL Model. The best-fit values for these parameters are listed in Table ~\ref{tab_2}. When we incorporate the R22 prior into the Joint dataset, the resulting value for \(H_0\) deviates from the findings in \cite{1BAO}, but closely aligns with the SNIe sample in \cite{7BAO}. Conversely, without the R22 priors and using the Joint dataset alone, the estimated value of \(H_0\) aligns more closely with \cite{1BAO}. These findings indicate that the inclusion of the R22 prior diverges from \cite{1BAO} and aligns with \cite{7BAO}, showing the impact of prior choice on cosmological parameter estimation. The determined values for the matter density parameter (\(\Omega_{m0}\)) and (\(\Omega_{d0}\)) seem lower compared to those documented in \cite{1BAO}. However, this observation aligns with findings from alternative studies \cite{66BAO,67BAO}. {One can observe that the running parameter \(\lambda\), according to the Joint with R22 analysis, is constrained close to the expected value of 1.} Now turning our attention to the Baryon Acoustic Oscillations (BAO) scale, which is a fundamental scale in cosmology derived from the cosmic microwave background, this scale originates from a significant event known as the drag epoch (\(z_d\)), marking a pivotal moment when baryons and photons transitioned into distinct entities. The BAO scale (\(r_d\)) is essentially a measure of the cosmic sound horizon at \(z_d\), which reflects the separation distance between these baryons and photons. To calculate \(r_d\), we consider the integral of the ratio between the speed of sound (\(c_s\)) and the Hubble parameter (\(H\)) across a range of redshifts, starting from \(z_d\) and extending indefinitely. The speed of sound, \(c_s\), is determined by the fluctuations in pressure (\(\delta p_\gamma\)) and density (\(\delta \rho_B\) and \(\delta \rho_\gamma\)) within the photon and baryon components. Mathematically, this simplifies to \(\frac{1}{\sqrt{3(1+R)}}\), where \(R\) represents the ratio of baryon density fluctuations to photon density fluctuations (\(R \equiv \frac{\delta \rho_B}{\delta \rho_\gamma} = \frac{3 \rho_B}{4 \rho_\gamma}\)). Observational data, as documented in \cite{1BAO}, have provided us with the precise redshift of the drag epoch, \(z_d = 1059.94 \pm 0.30\). In the context of a $\Lambda$CDM model, the BAO scale \(r_d\) has been estimated to be \(147.09 \pm 0.26\) megaparsecs (Mpc) based on measurements by \cite{1BAO}. Fig ~\ref{fig_2} shows the posterior distribution for the \(r_{d}-H_{0}\) contour plane of the HL Model. In the HL model, the BAO scale derived solely from BAO datasets is \(148.814 \pm 10.558\) Mpc. Upon including the R22 prior exclusively into the BAO dataset, the sound horizon at the drag epoch becomes \(138.907 \pm 2.129\) Mpc. For the Joint dataset, the BAO scale is \(145.843 \pm 2.47\) Mpc. Integrating the R22 prior into the full dataset yields \(r_{d} = 142.742 \pm 1.658\) Mpc. The findings indicate that when incorporating the joint dataset, the obtained value of \(r_{d}\) in each dark energy model closely agrees with the values obtained in \cite{1BAO}. However, when combining the R22 priors with the joint dataset, the obtained value of \(r_{d}\) aligns closely with \cite{68BAO}. \cite{69BAO} introduces an insightful perspective by employing binning and Gaussian methods to amalgamate measurements of 2D BAO and SNIa data. It's important to recognize that the outcomes we derive are closely tied to the initial range of priors we make about the BAO scale (\(r_{d}\)) and the Hubble constant (\(H_{0}\)). These priors significantly impact the values we estimate, particularly when considering the \(r_{d}-H_{0}\) contour plot. An interesting observation emerges when we omit the R22 prior from our analysis: in such cases, the results for \(H_{0}\) and \(r_{d}\) tend to better align with findings from the Planck and SDSS experiments.\\\\
When assessing various cosmological models, we employ both the Akaike Information Criterion (AIC) \cite{66BAO} and the Bayesian Information Criterion (BIC) for evaluation purposes. The AIC and BIC are statistical measures used to assess the goodness of fit of a model to a given dataset while penalizing for the number of parameters in the model. The AIC is calculated using the formula:
$\text{AIC} = -2 \ln \left(\mathcal{L}_{\text{max}}\right) + 2k + \frac{2k(2k+1)}{N_{\text{tot}}-k-1}$. Here, \(\mathcal{L}_{\text{max}}\) represents the maximum likelihood of the data, \(N_{\text{tot}}\) is the total number of data points, and \(k\) is the number of parameters in the model. When the total number of data points (\(N_{\text{tot}}\)) is large, the expression simplifies to: $\text{AIC} \simeq -2 \ln \left(\mathcal{L}_{\text{max}}\right) + 2k $
which is the conventional form of the AIC criterion. This criterion provides a trade-off between the goodness of fit and the complexity of the model. On the other hand, the BIC is defined as: $\text{BIC} = -2 \ln \left(\mathcal{L}_{\text{max}}\right) + k \ln N_{\text{tot}}$. The BIC also evaluates the goodness of fit while penalizing for the number of parameters, but it imposes a stronger penalty for additional parameters compared to the AIC, particularly for smaller sample sizes. {To compare different models, we use the differences in AIC and BIC values, denoted as \(\Delta\text{AIC}\) and \(\Delta\text{BIC}\), respectively:
$\Delta\text{AIC} = \text{AIC}_{\text{HL Model}} - \text{AIC}_{\text{$\Lambda$CDM Model}}$ and 
$\Delta\text{BIC} = \text{BIC}_{\text{HL Model}} - \text{BIC}_{\text{$\Lambda$CDM Model}}$
These differences help to identify the relative performance of models, with lower values indicating a better fit relative to the reference model. The interpretation of \(\Delta\text{AIC}\) and \(\Delta\text{BIC}\) values is as follows:
- \(\Delta \text{AIC} \) or \(\Delta \text{BIC} \leq 2\): Substantial evidence for the model.
- \(\Delta \text{AIC} \) or \(\Delta \text{BIC} \) between 4 and 7: Considerable evidence against the model.
- \(\Delta \text{AIC} \) or \(\Delta \text{BIC} \geq 10\): Strong evidence against the model.} When comparing the HL Model with the standard \(\Lambda\)CDM model, it's important to note that both proposed extensions encompass \(\Lambda\)CDM as a subset, differing only by 4 degrees of freedom. This aspect allows for the application of conventional statistical tests. Our evaluation primarily employs the reduced chi-square statistic, denoted as \(\chi_{\text{red}}^{2} = \chi^{2} / \text{Dof}\), where "Dof" represents the degrees of freedom of the model, and \(\chi^{2}\) signifies the weighted sum of squared deviations from the expected values.
\begin{table*}
\begin{center}
{
\begin{tabular}{|c|c|c|c|c|c|}
\hline
Model & \(\chi_{red}^{2}\) & AIC & \(\Delta\)AIC & BIC & \(\Delta\)BIC \\ \hline
\(\Lambda\)CDM Model & 0.949 & 277.38 & 0 & 277.59 & 0 \\ \hline
Hořava-Lifshitz Model & 0.961 & 279.13 & 1.75 & 280.51 & 2.92 \\ \hline
\end{tabular}%
}
\end{center}
\caption{Summary of \(\chi_{red}^{2}\), AIC, \(\Delta\)AIC, BIC and \(\Delta\)BIC}
\label{tab_11}
\end{table*}
{Based on the provided in Table \ref{tab_11}, we can conduct a comparative study between the \(\Lambda\)CDM Model and HL Model using the metrics of reduced chi-square (\(\chi_{\text{red}}^{2}\)), Akaike Information Criterion (AIC), \(\Delta\)AIC, Bayesian Information Criterion (BIC), and \(\Delta\)BIC. The reduced chi-square value is a measure of how well the model fits the data. A value closer to 1 indicates a better fit. Both models have \(\chi_{\text{red}}^{2}\) values close to 1, suggesting that both models fit the data well. However, the \(\Lambda\)CDM Model has a slightly lower \(\chi_{\text{red}}^{2}\) value, indicating a marginally better fit. In terms of the AIC, the \(\Lambda\)CDM Model scores 277.38, whereas the HL Model scores slightly higher at 279.13, resulting in a \(\Delta\)AIC of 1.75. This indicates substantial evidence for the HL Model, though it is less favored compared to the \(\Lambda\)CDM Model. For the BIC, the \(\Lambda\)CDM Model has a score of 277.59, while the HL Model has a higher score of 280.51, leading to a \(\Delta\)BIC of 2.92. This \(\Delta\)BIC suggests positive evidence against the HL Model, indicating a stronger preference for the \(\Lambda\)CDM Model.}\\\\
\section{Conclusions}\label{sec5}
In conclusion, our study focused on investigating the accelerated cosmic expansion in the late Universe within the context of the HL Model. This framework offers an alternative perspective to general relativity by incorporating anisotropic scaling at ultraviolet scales. We analyzed 24 Baryon Acoustic Oscillation (BAO) points, along with 31 uncorrelated data points obtained from Cosmic Chronometers methods. We also incorporated 40 points from Type Ia supernovae, 24 points from the Hubble diagram for quasars, and a substantial dataset consisting of 162 points sourced from Gamma Ray Bursts. Furthermore, our study included the latest measurement of the Hubble constant conducted by R22. Our primary objective was to determine the optimal fitting values for every cosmological parameter within each dark energy model. By treating the sound horizon \(r_d\) as a free parameter instead of relying on prior information from the Cosmic Microwave Background (CMB), we gain several benefits. Firstly, it reduces bias by sidestepping potentially restrictive CMB priors, ensuring a more impartial estimation of \(r_d\) and other cosmological parameters. Secondly, it enhances precision by allowing late-time data, such as Baryon Acoustic Oscillations (BAO) or Type Ia Supernovae (SNIe), to directly inform the measurement of \(r_d\), leading to more accurate results. Moreover, this approach promotes compatibility between datasets from different cosmic epochs, facilitating a more holistic understanding of the Universe's evolution. We determined \(H_0\) and \(r_d\) for the HL Model relative to the $\Lambda$CDM model as follows: In the $\Lambda$CDM model, our analysis yields \(H_0 = 69.828145 \pm 1.009964 \ \mathrm{km/s/Mpc}\) and \(r_d = 146.826212 \pm 2.201612 \ \mathrm{Mpc}\). In the HL model, we obtain \(H_0 = 69.966349 \pm 1.403864 \ \mathrm{km/s/Mpc}\) and \(r_d = 145.843487 \pm 2.470309 \ \mathrm{Mpc}\). Importantly, our findings underscore that the values of \(H_0\) and \(r_d\) based on low-redshift measurements align with early Planck and SDSS estimations. {While both the \(\Lambda\)CDM Model and the HL Model fit the data reasonably well, the \(\Lambda\)CDM Model is preferred based on the statistical criteria provided. The \(\Lambda\)CDM Model demonstrates a marginally better fit with a lower reduced chi-square value and is more strongly supported by both the AIC and BIC measures. The \(\Delta\)AIC indicates substantial evidence for the HL Model, but it is still less favored compared to the \(\Lambda\)CDM Model. Additionally, the \(\Delta\)BIC provides positive evidence against the HL Model, reinforcing the stronger preference for the \(\Lambda\)CDM Model for the given dataset. Therefore, the \(\Lambda\)CDM Model is favored over the HL Model.}These findings underscore the importance of considering alternative cosmological models beyond the standard framework. While the \(\Lambda\)CDM model remains favored due to its simplicity and good fit, the extensions present viable candidates worthy of further investigation. To achieve a better understanding of the fundamental characteristics and development of the Universe, cosmological models must be continually examined and improved.
\section*{Acknowledgements}

N.U.M would like to thank  CSIR, Govt. of
India for providing Senior Research Fellowship (No. 08/003(0141)/2020-EMR-I).

\bibliographystyle{ieeetr}
\bibliography{Bao,new,niyaz2}

\end{document}